\documentclass[aps,prd,floatfix,superscriptaddress,twocolumn,showpacs]{revtex4}
\usepackage{epsfig}
\usepackage{amsmath}
\usepackage{graphicx,psfrag}
\usepackage{dcolumn}
\usepackage{bm}
\usepackage{slashed}
\usepackage{xcolor}
\usepackage{maybemath}

\newcommand{\beq}{\begin{equation}}
\newcommand{\eeq}{\end{equation}}
\newcommand{\hf} {\frac{1}{2}}

\newcommand{\nn}{\nonumber\\}

\newcommand{\be}{\begin{equation}}
\newcommand{\ee}{\end{equation}}
\newcommand{\bea}{\begin{eqnarray}}
\newcommand{\eea}{\end{eqnarray}}

\newcommand\fig[1]     {Fig.\,{\ref{#1}}}

\newcommand\app[1]     {Appendix~\ref{#1}}

\def\Tr{{\rm Tr}}

\def\eq#1{(\ref{#1})}
\def\s0#1#2{\mbox{\small{$ \frac{#1}{#2} $}}}
\def\0#1#2{\frac{#1}{#2}}

\def\mr#1{{\mathrm{#1}}}

\begin{document}
 
\title{Renormalisation of non-differentiable potentials}

\author{J.~Alexandre}
\affiliation{Department of Physics, King's College London, WC2R 2LS, UK}

\author{N.~Defenu}
\affiliation{Institut f\"ur Theoretische Physik, 
Universit\"at Heidelberg, D-69120 Heidelberg, Germany}

\author{G.~Grigolia}
\affiliation{University of Debrecen, P.O.Box 105, H-4010 Debrecen, Hungary}
\affiliation{University of Trento, Department of Physics, Via Sommarive 14, 38123 Povo, Trento, Italy}

\author{I.~G.~M\'ari\'an}
\affiliation{Atomki, P.O.~Box 51, H--4001 Debrecen, Hungary} 
\affiliation{University of Debrecen, P.O.Box 105, H-4010 Debrecen, Hungary}
\affiliation{MTA--DE Particle Physics Research Group, 
P.O.~Box 51, H--4001 Debrecen, Hungary}

\author{D.~Mdinaradze}
\affiliation{University of Debrecen, P.O.Box 105, H-4010 Debrecen, Hungary}

\author{A.~Trombettoni}
\affiliation{Department of Physics, University of Trieste, Strada Costiera 11, I-34151 Trieste, Italy}
\affiliation{CNR-IOM DEMOCRITOS Simulation Center, 
Via Bonomea 265, I-34136 Trieste, Italy}

\author{Y.~Turovtsi-Shiutev}
\affiliation{Uzhhorod National University,
14, Universytetska str., Uzhgorod, 88000, Ukraine}

\author{I.~N\'andori}
\affiliation{University of Debrecen, P.O.Box 105, 
H-4010 Debrecen, Hungary}
\affiliation{MTA--DE Particle Physics Research Group, 
P.O.~Box 51, H--4001 Debrecen, Hungary}
\affiliation{Atomki, P.O.~Box 51, H--4001 Debrecen, Hungary} 

\begin{abstract} 
Non-differentiable potentials, such as the $V$-shaped (linear) potential, appear in various areas of physics. 
For example, the effective action for branons in the framework of the brane world scenario 
contains a Liouville-type interaction, i.e., an exponential of the $V$-shaped function. Another example is 
coming from particle physics when the standard model Higgs potential is replaced by a periodic 
self-interaction of an N-component scalar field which depends on the length, thus it is $O(N)$ symmetric.
We first compare classical and quantum dynamics near non-analytic points and discuss in this context
the role of quantum fluctuations. We then study the renormalisation of such potentials, focusing on the
Exact Wilsonian Renormalisation approach, and we discuss how quantum fluctuations smoothen the 
bare singularity of the potential. Applications of these results to the non-differentiable effective 
branon potential and to the $O(N)$ models when the spatial dimension is varied and to the $O(N)$ 
extension of the sine-Gordon model in (1+1) dimensions are presented.
\end{abstract}


\maketitle 

\noindent KCL-PH-TH/2021-77


\vspace{1cm}

\section{Introduction}

Non-differentiable potentials appear in various areas of physics. In classical physics they can emerge 
from the occurrence of particular constraints in certain points of the space and are typically associated 
to the presence of discontinuous forces. The latter are in turn associated to non-analytic points of the 
potential energy $V$, an example being a potential of the form $V(x) \propto |x|$, where the discontinuity 
takes place in $x=0$. Since the quantum properties of the models with such potential can be studied by 
Renormalisation Group (RG) relating via the flow equations their properties at different length scales, a 
natural question is to study how these potentials are renormalized and how in the flow equations the 
non-analytic behaviour is evolving.

As a first motivation for the study of their quantization and their properties in the context of RG, 
we observe that in the framework of models of large extra dimensions, in particular in the so-called 
brane world scenario (BWS), elementary particles except for the graviton are localized on 
(3+1)-dimensional branes. Although experimental tests from the Large Hadron Collider severely 
constrain theories of large extra dimensions, the BWS served as one of the simplest extensions 
of the Standard Model. The brane fluctuations of the BWS in a 5th dimensional bulk lead to a low 
energy  effective four-dimensional theory, where branons (representing quanta of the brane 
fluctuations) are described by a scalar field living on a flat brane (for references on branon studies 
see e.g., \cite{Kugo,Bando,Dobado,Cembranos,branon_stabilization,Burnier,branon_dressing}). 
Assuming the brane centered on a Randall-Sundrum \cite{RS} warp factor, the effective 
branon theory in (3+1) dimensions involves then the absolute value of the branon field, leading 
to a non-differentiable potential and wave function renormalization, 
\beq
\label{Sbranon}
S_{\mr{Branon}} = \int d^4x\left( \frac{e^{-2a\vert\phi \vert}}{2}\partial^\mu\phi\partial_\mu\phi 
- f^4e^{-4a\vert\phi \vert}\right),
\eeq
where the Liouville-type terms in the potential and in the wave function renormalization 
depend on the absolute value of the scalar field, i.e., it is non-differentiable, as further discussed in
\app{app1}.

Another major example for non-differentiable potentials is coming from particle physics when the standard 
model quartic and quadratic Higgs potential is replaced by a periodic one, with the Higgs part given by the 
following Euclidean action \cite{sg_msg_higgs_1,sg_msg_higgs_2}:
\begin{equation}
\label{o(N)_sg}
S_{\mr{Higgs}} = \int d^4 x \,\Big( \hf \partial_{\mu} \boldsymbol{\phi} \partial_{\mu} \boldsymbol{\phi} 
+ u  \cos(\beta\sqrt{\boldsymbol{\phi} \cdot \boldsymbol{\phi}}) \Big)
\end{equation}
where $\boldsymbol{\phi}(x)$ is an $N$-component scalar field, see \app{app2}.

A third example is a non-differentiable potential of the form
\be
V(\phi)=V_0\Big(1-[A+\theta(\phi-\phi_0)\Delta A](\phi-\phi_0)\Big)~,
\ee
used in an inflationary scenario \cite{slowroll1,slowroll2}, where the field $\phi$ sees a discontinuous slope
at $\phi_0$, in order to obtain a better data fit to the power spectrum for density perturbations during inflation.

These examples show how discontinuities can emerge in different effective models, and we note that we 
consider here interactions which are not differentiable in the field, unlike involving a background field which 
is not differentiable with respect to space time coordinates \cite{Bordag:2004rx}.

Exact solutions of the one-dimensional Schr\"odinger equation with non-differentiable linear $V$-shaped 
potential $V(x)=u\vert x\vert$, and exponential potentials of the form $V(x)=-g^2 \exp(-\vert x \vert)$ 
or $V(x)=g^2 \exp(2 \vert x \vert)$, have already been discussed in literature, see \cite{sakurai} and  
\cite{qm_exp_pot,qm_exp_analytic}, respectively. The exponential potentials are simplified "toy" versions 
of the more general branon action discussed above. The exact solutions for these non-relativistic, 
non-differentiable potentials suggest that their non-differentiable nature makes no difficulties if one 
considers their quantised theory compared to the classical description. However, interesting questions are 
related to show whether one might expect a similar result for higher dimensions ($d > 1$), if one considers 
the quantum theory corresponding to these non-analytic potentials; and whether their non-differentiable 
nature could create any difficulties in their renormalization. To answer these questions is the main goal of 
the present paper. One can expect on general ground that quantum fluctuations may provide a mechanism 
to soften/round the non-analyticities present in the bare models. We plan to use a variety of methods, including 
renormalization group based calculations, to clarify qualitatively and quantitatively such point.

To present a defined path for a clear formulation of the problem, we first illustratively give 
a hint of the classical dynamics around non-analytic point and discuss how the dynamics of a wave function 
is modified in the quantum case. This will give an hint on the understanding of how quantum fluctuations 
smoothen the behaviour of a system near such non-analytic points, a result that we confirm via RG techniques 
extending it to $O(N)$ field theories. The advantage of the RG formulation we are going to present is that one 
can then study different dimensions within a single formalism.

As a further application of the present formalism, we will study periodic extension of $O(N)$ models. It is well 
known that for $N=1$ the non-differentiable periodic model \eq{o(N)_sg}, known as the sine-Gordon model, 
has a topological phase transition.  We will  study the phase structure of \eq{o(N)_sg} for $N>1$ and compare 
the obtained results with large-$N$ findings.

Summarizing, our goal in this work is two-fold: from one side we aim at studying whether the non-analitycal 
behavior of the potential such as the one in  \eq{Sbranon} conflicts with its quantisation and renormalization; 
from the other, we investigate whether $O(N)$ sine-Gordon has a topological phase transition for $N>1$.

\section{Non-analytic potentials in classical mechanics}
\label{classical}
Let us first discuss the physics of non-differential potentials  in the framework of classical 
(non-relativistic and not quantised) mechanics. We consider two different non-differentiable
potentials in $d=1$ dimensions:
\bea
\label{abs}
V^{1}(x) &=& u \vert x \vert, \\
\label{exp_abs}
V^{2}(x) &=& u \exp(a\vert x \vert).
\eea
In particular we are interested in the solution of the equation of motion for the potentials 
\eq{abs} and \eq{exp_abs}.  In order to investigate the impact of the non-differentiable nature 
of the potential in $x=0$ on the classical solution one can use its regularised form. This can be 
achieved by introducing a regularised form of the potentials, such as
\bea
\label{abs_reg}
V^{1}_{\mr{reg}}(x) &=& u \sqrt{r^2 + x^2}, \\
\label{exp_abs_reg}
V^{2}_{\mr{reg}}(x) &=& u \exp(a \sqrt{r^2 + x^2}).
\eea
which are differentiable in $x=0$, with the limit $r\to 0$ recovering the original potentials
\eq{abs} and \eq{exp_abs}.

\begin{figure}[ht]
\begin{center}
\includegraphics[width=0.8\linewidth]{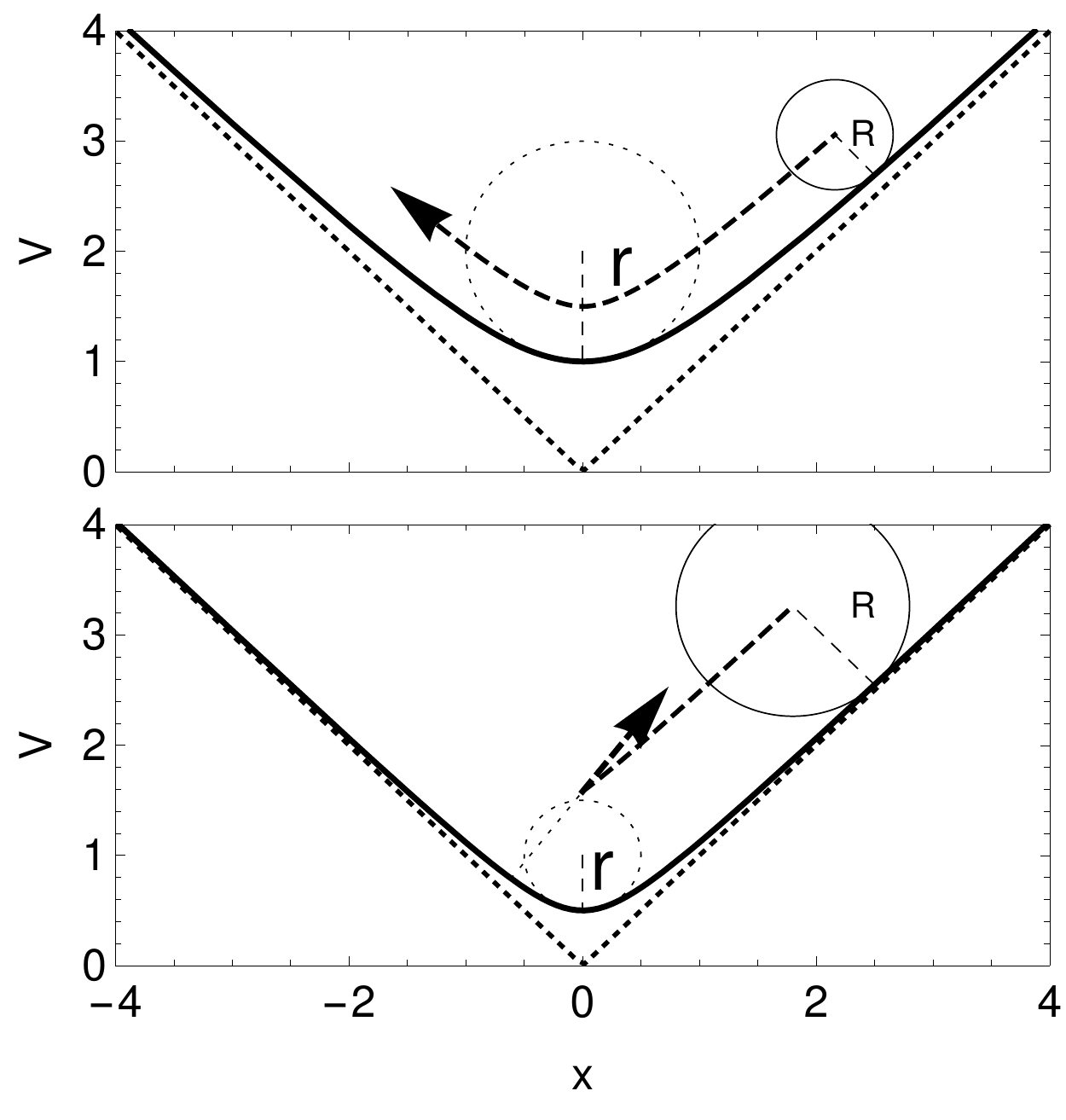}
\caption{
Classical motion of a ball with a finite radius $R$ in the regularized potential \eq{abs_reg} with $u=1$, 
which is characterized by the radius of curvature $r$ at $x=0$, visualized by the dotted osculating circle. 
The motion of the ball depends on the ratio $r/R$. If $r/R \gg 1$, then the ball can smoothly roll to the 
other side (top panel), while if $r/R \ll 1$, then it hits the other side of the potential (considering it as a 
hard surface) and might bounce up (bottom panel).
} 
\label{fig1}
\end{center}
\end{figure}

If one considers the classical motion of a ball with a finite radius $R$ in the regularised (1-dimensional) 
potential \eq{abs_reg} with $u=1$ one finds two different cases with two different type of motions. 
For large values of $r$, when $r/R \gg 1$, the ball can smoothly roll or slide over to the other side of 
the potential, since the largest radius of curvature is $r$ at $x=0$. In this case the ball oscillates on both 
sides of the potential hill, see the top panel of \fig{fig1}.
When $r/R \ll 1$ the ball cannot smoothly roll or slide over to the other side, and a different motion occurs. 
For example considering the potential as a hard surface in a constant gravitational field, the ball might 
bounce up, see the bottom panel of \fig{fig1}. Thus, the non-differentiable nature of the potential does 
matter for extended objects and in this case the limit $r\to 0$ is non-analytic. However, if one considers
the force derived from the regularised potential acting on a point-like particle, the limit $r\to 0$ 
is analytic and one finds oscillating motion on both sides even for $r=0$. 
%
%
%

One can think of many ways to regularize a non-analytic potential other than the examples presented in 
Eq.~\eq{abs_reg} and \eq{exp_abs_reg}. One of the possibilities is to simply replace the original non-analytic 
potential in an interval $(-r,r)$ around the point of non-analycity with a differentiable function $f(x)$, while 
outside this interval the potential remains unchanged. 
Here we give two examples of this kind of regularization in the case of a one dimensional $u|x|$ type potential.

Similar to the initial potential we require $f(x)$ to be an even function $f(-x)=f(x)$.
For the regularized potential to be continuous and differentiable the relations 
$f(r)=V(r)$, $f'(r)=V'(r)$ must hold as well. The simplest function which can fulfill all 
the given requirements is a quadratic polynomial, thus the regularized potential takes the form
\beq
\label{quadratic_reg}
V^{3}(x)=\begin{cases}
u|x|, & |x| \geq r \\
\frac{u}{2r} x^2 + \frac{ur}{2}, & |x|<r \, .
\end{cases}
\eeq
This model essentially interpolates between the non-analytic linear potential and the potential of a
harmonic oscillator.

With this regularization method it is also possible to require the continuity of higher order derivatives using 
higher order polynomials. In the case of the linear type potential the higher order derivatives are continuous 
(up to the second order) if the replacement function is a quartic polynomial
\beq
\label{quartic_reg}
V^{4}(x)=\begin{cases}
u|x|, & |x| \geq r
\\
\frac{-u}{8r^3} x^4 + \frac{3u}{4r} x^2 + \frac{3ur}{8} , & |x|<r \, .
\end{cases}
\eeq

In the next section 
we study these potentials and observe how quantum physics modifies the classical results.

\section{Non-analytic potentials in quantum mechanics}
As a next step one should consider the quantum version of the 1-dimensional problems considered 
in the previous Section. Let us note, that in $d=1$ dimensions the quantum field theory for a single 
scalar field reduces to a simple 1-dimensional quantum mechanical problem, since the field variable 
can be replaced by the position of a particle $\phi \to x$. It can be shown that due to the fact that the 
potential is non-differentiable only at a single point (at $x=0$), its quantum theory does not suffer from 
the non-analytic behaviour of the potential. Thus, the non-differentiable nature of the potential does 
not matter in the quantum theory if it is non-differentiable only at a single point. We want also to discuss 
if and how the classical regimes represented in \fig{fig1} are present in the quantum case. We will also 
discuss the ground state energy of the quantum mechanical problem, and similar to the classical case, 
see whether the limit $r\to 0$ is non-analytic.

For this reason let us examine the eigeinvalues and eigenvectors of the one-dimensional 
Hamiltonian defined as
\beq
\label{qm_ham}
H=-\frac{d^2}{dx^2}+V(x)
\eeq
omitting the prefactor of the derivative for simplicity ($\hbar=m=1$).
For both potentials \eq{abs}, \eq{exp_abs} there are analytic solutions. To deal with the absolute 
value the parameter space is restricted to $x\geq0$. The potentials have $Z_2$ symmetry therefore 
the solutions must have even or odd parity with either $\psi(x=0)=0$ or $\psi'(x=0)=0$. Using these 
conditions one can extend the solutions for the negative region. 

The eigenvector for the linear potential can be described by the Airy function $\psi \propto 
Ai[u^{1/3} (x-E/u)]$ \cite{sakurai}. The energy spectrum is determined by the zeros of the Airy 
function and the zeros of its first derivative, thus the ground state energy is $E_0\approx 1.019 u^{2/3}$.
The eigenvector for the exponential potential \eq{exp_abs} can be constructed with Bessel functions 
\cite{qm_exp_analytic}, and the energy state is, again, given by the zeros of the Bessel function and 
the zeros of its derivative. Therefore the ground state for $a=2$ (and for $u=1$) is approximately 
$E_0\approx 3.676$.

\begin{figure}[ht]
\begin{center}
\includegraphics[width=0.8\linewidth]{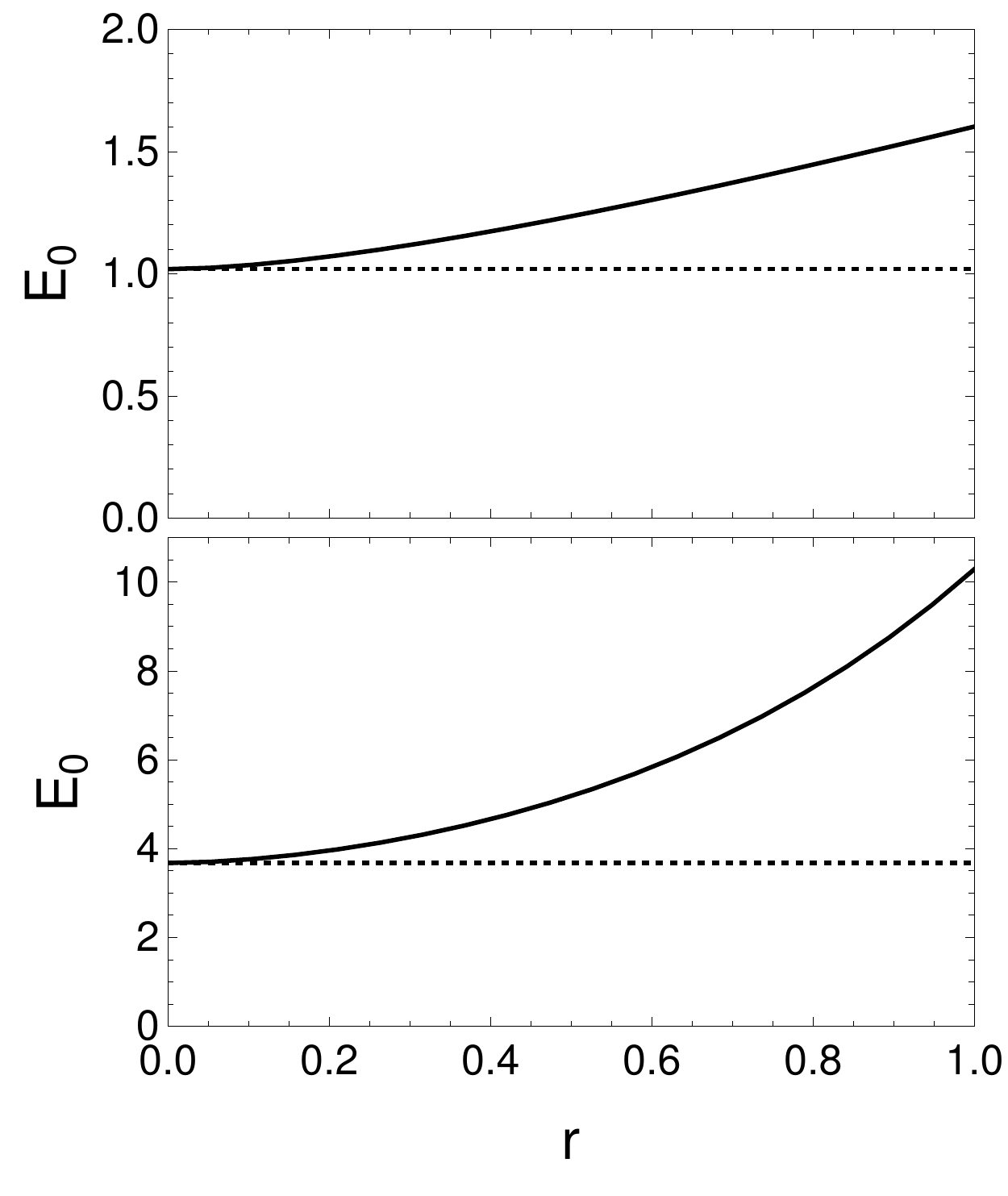}
\caption{
Dependence of the ground state energy $E_0$ on the regularization parameter $r$ calculated for 
the quantum mechanical system with linear potential \eq{abs_reg} with $u=1$ (top) and exponential 
potential \eq{exp_abs_reg} with $a=2$ and $u=1$ (bottom). The dotted lines indicate the results 
obtained from the analytic solutions for the original ($r\to 0$) potentials \eq{abs} and \eq{exp_abs}.
} 
\label{fig2}
\end{center}
\end{figure}

\begin{figure}[ht]
\begin{center}
\includegraphics[width=0.8\linewidth]{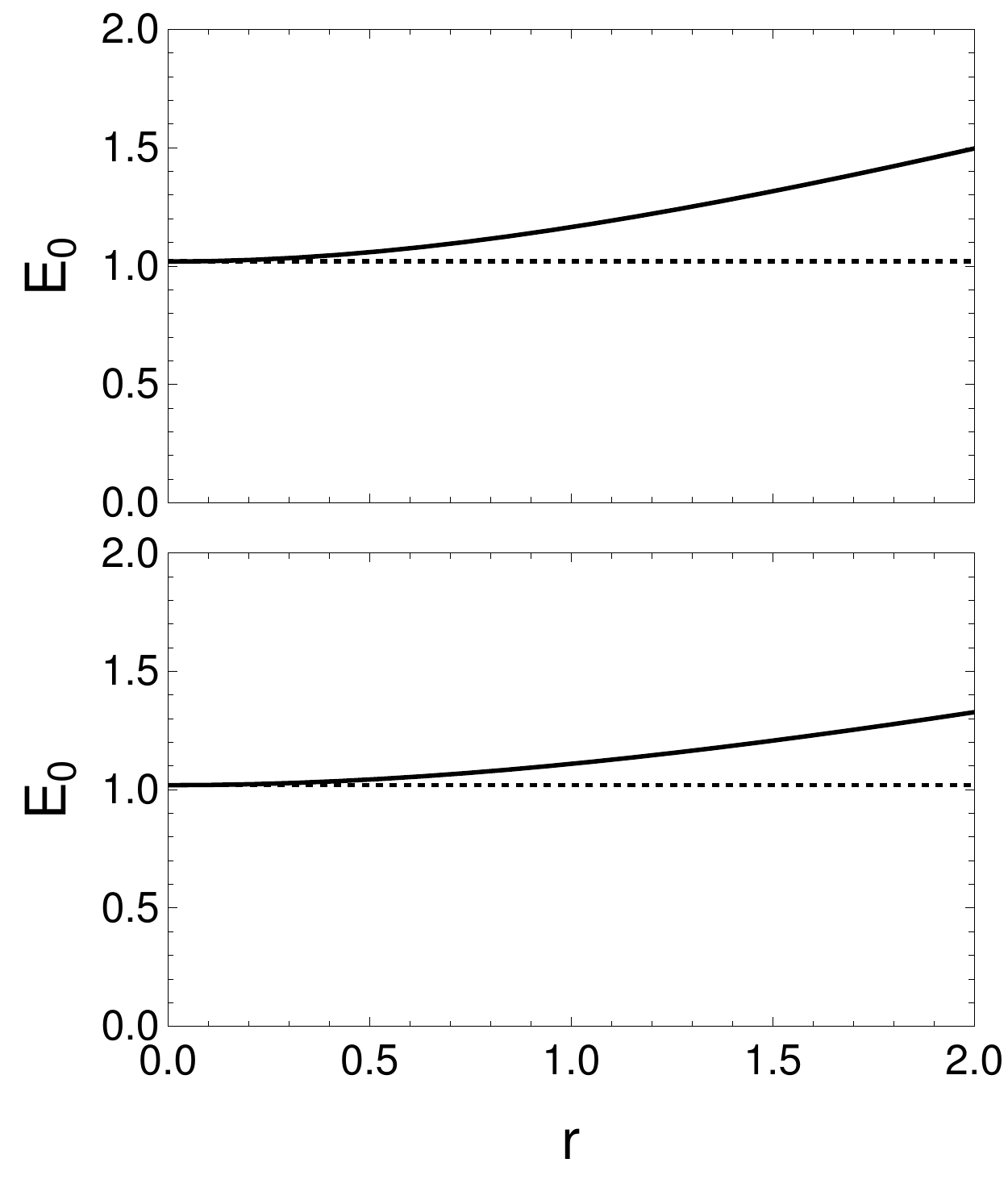}
\caption{
Dependence of $E_0$ on $r$ calculated for the quantum mechanical system with linear potential 
regularized using a quadratic polynomial \eq{quadratic_reg} with $u=1$ (top) and a quartic polynomial 
\eq{quartic_reg} with $u=1$ (bottom). The dotted lines indicate the results obtained from the analytic 
solutions for the original ($r\to 0$) potential \eq{abs}.
} 
\label{fig3}
\end{center}
\end{figure}

Now the question is whether the regularized potentials \eq{abs_reg}, \eq{exp_abs_reg} and 
\eq{quadratic_reg}, \eq{quartic_reg} recover the above analytic solutions in the $r\to 0$ limit. \fig{fig2} 
shows the numerical solution for the regularized potentials \eq{abs_reg} and \eq{exp_abs_reg}. As $r$ 
increases the ground state energy as a function of $r$ tends to a linear or exponential function 
respectively, which follows from examining the minimum value of the potentials. 
%
%
%
It is also clear that in the $r \to 0$ the ground state energy is recovered continuously. 
Similarly, \fig{fig3} shows the numerical solution for the regularized potentials \eq{quadratic_reg} 
and \eq{quartic_reg}. As expected for the linear potential regularized with a quadratic function 
\eq{quadratic_reg}, the ground state energy of the model interpolates between the ground state 
energy of the non-analytic linear potential and the potential of a harmonic oscillator. As \fig{fig3} 
shows, in the $r \to 0$ limit, the result obtained from the analytic solution for the linear potential is 
recovered, while for $r \gg 1$ the result is a close approximation of a harmonic oscillator, since the 
$|x| \geq r$ region can be neglected. \fig{fig3} also shows that the requirement of the continuity of 
higher order derivatives, which is achieved by the regularization of the linear potential with a quartic 
function \eq{quartic_reg}, only slightly modifies the results.
%
%
Importantly, in all cases the known solution in the ${r \to 0}$ limit is continuously recovered, thus the 
limit is indeed analytic.

\begin{figure}[ht]
\begin{center}
\includegraphics[width=0.8\linewidth]{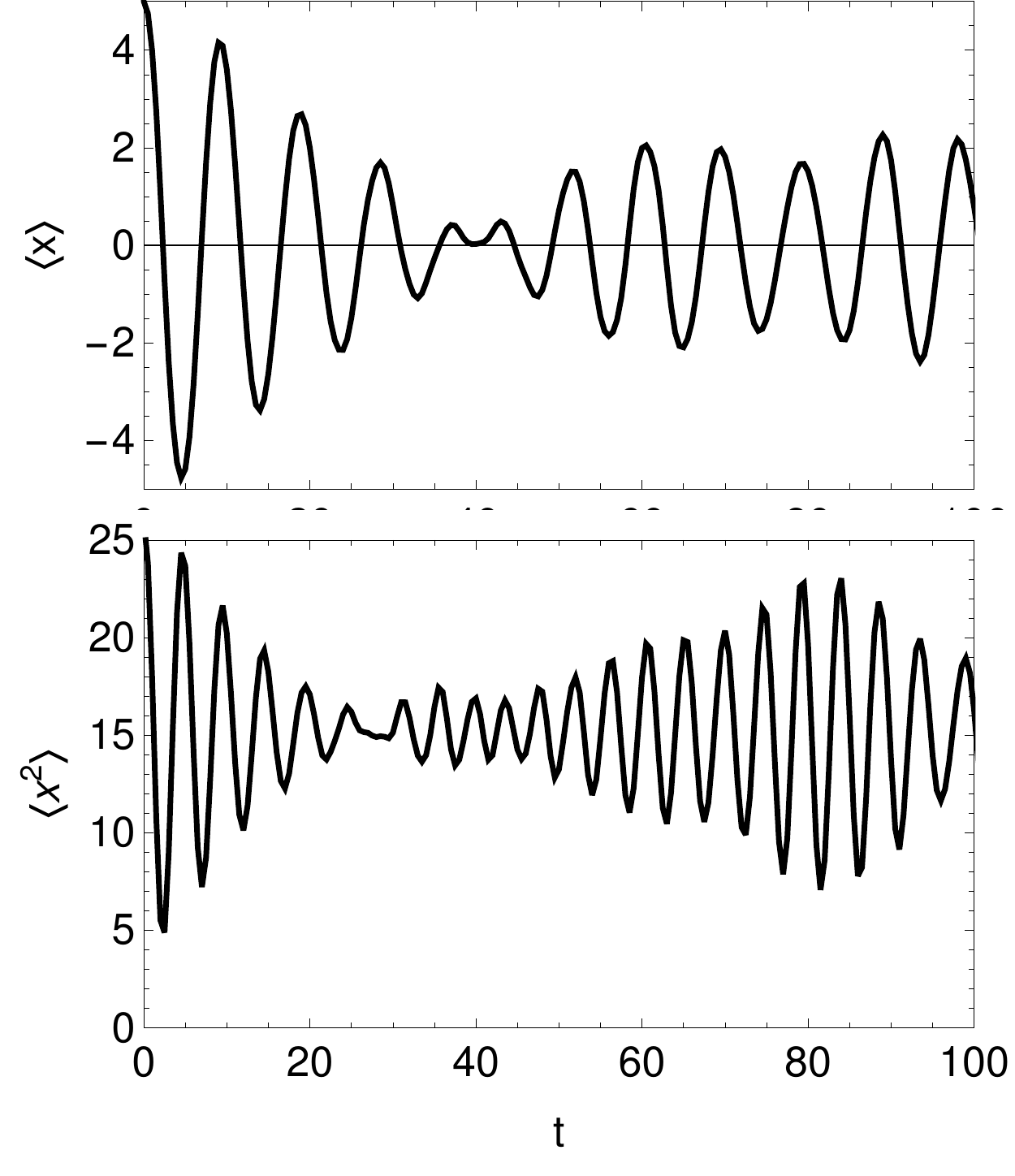}
\caption{ 
Expectation value of $x$ and $x^2$ of a wave packet as a function of time moving in a V shaped potential \eq{abs} 
with $u=1$. The initial wave packet is Gaussian \eq{initial_wavepacket} with $x_0=5$ and $\sigma_0=1$ .
} 
\label{fig4}
\end{center}
\end{figure}

Let us also examine the dynamics of this quantum mechanical problem, solving the Schr\"odinger equation
\beq
\label{sch_eq}
i \partial_t \psi(x,t)=H \psi(x,t),
\eeq
where the Hamiltonian is given by \eq{qm_ham}.
The question is what happens to a Gaussian wave packet in a non-differentiable, V shaped potential \eq{abs}.
The initial condition for the wave packet is given as 
\beq
\label{initial_wavepacket}
\psi(x,0)= \frac{1}{(2 \pi  \sigma^2_0)^{4}} \exp\left[- \frac{(x - x_0)^2}{4 \sigma_0^2}\right].
\eeq
In order to solve \eq{sch_eq}, one can as usual expand the initial Gaussian wave packet \eq{initial_wavepacket} 
in terms of the eigenfunctions $\phi_n$ of the Hamiltonian by calculating the overlaps as
\beq
\psi(x,0)=\sum_n c_n \phi_n(x), \hskip0.3cm 
c_n=\int dx \, \psi^*(x,0) \phi_n(x).
\eeq
As discussed above, in the case of a linear potential the eigenfunctions are Airy functions, thus for the 
potential \eq{abs} we have
\beq
\phi_n=A_n \, \mbox{sign}(x)^{n+1} \, Ai \left[ u^{1/3} (|x|+ E_n /u) \right] ,
\eeq
where $A_n$ is the normalization factor and $E_n$ is the energy spectrum, which as stated before, can 
determined by the zeros of the Airy function and its derivative. Then, applying the time evolution operator 
$e^{-iH t}$ is straigthforward, since one has to simply multiply each eigenfunction with the corresponding 
factor $e^{-iE_n t}$. This yields the desired solution of the wavefunction at time $t$,
\beq
\psi(x,t)=\sum_n e^{-iE_n t}c_n \phi_n(x).
\eeq
To see how this wave packet evolves, the expectation value of $x$ and $x^2$ is plotted as a function of time in \fig{fig4}.
%
%
%
The consistency of the solutions has been checked also by directly solving the Schr\"odinger equation 
numerically. As \fig{fig4} shows, the wave packet is oscillating on both the positive and negative region of 
the potential. The wavefunction remains smooth and well defined during its evolution and of course quite 
different from the harmonic $x^2$ potential.  When comparing \fig{fig4} with \fig{fig1}, one sees that the 
dynamics of the wave packet is clearly more the counterpart of the oscillating dynamics in the top part of 
\fig{fig1}. Since the bouncing dynamics in the bottom part of \fig{fig1} is due to the effect of the singularity in 
$x=0$ seen by the ball as a wall there present, we see an example of how the quantum fluctuations prevents 
the bouncing of the wave packet and the effective rounding of the effective potential seen by it. Another way 
to understand this result is by observing that the motion of the center of motion is given by Ehrenfest's theorem, 
in which the center of mass is the counterpart of the a point-like particle, being therefore in the situation 
$r/R \gg 1$ case of the top part of \fig{fig1}. We conclude that in quantum mechanics the ground state energy 
and the dynamics of a wave packet can be determined for the studied non-differentiable potentials and the 
quantum fluctuations effectively smoothen the potential near the non-analytic point.

One might expect a similar result for higher dimensions ($d>1$), i.e., if one considers 
the quantum theory of the non-analytic potentials, their non-differentiable nature might not create 
any difficulties. In other words, the renormalization group transformations can be performed by for 
example the grid-method and the single point where the potential is non-differentiable can be 
handled. Our goal in the next sections is to perform this RG study.

\section{Functional RG equation for single scalar models}

\subsection{Differentiability and the Wilsonian effective potential}

We show here that, even for a non-differentiable microscopic potential, the Wilsonian effective 
potential $V_{eff}$ should be differentiable. The latter can be defined by integrating over all Fourier 
components of the field, except for the constant mode, which is fixed to some value $\phi_0$ 
\be 
e^{i{\cal V}V_{eff}(\phi_0)}=\int{\cal D}[\phi] e^{iS[\phi]}~\delta\Big(\int d^dx(\phi-\phi_0)\Big)~,
\ee 
where ${\cal V}$ is the spacetime volume, which needs to be fixed in order to regularise the path integral.
Note the use of the Minkowski metric, which is essential for the following argument to be valid. 

We write the Dirac distribution as the Fourier transform of an exponential, such that
\bea 
e^{i{\cal V}V_{eff}(\phi_0)}&=&\int{\cal D}[\phi] e^{iS[\phi]}~\int dj~e^{ij\int d^dx(\phi-\phi_0)}\\
&=&\int dj~e^{-ij{\cal V}\phi_0}~Z[j]~,\nonumber
\eea
where $Z[j]$ is the partition function for the constant source $j$. 
The right-hand side integral is differentiable in $\phi_0$, hence we expect the Wilsonian effective potential 
$V_{eff}$ to be differentiable. In the following subsections we study a specific example which illustrates this feature.

\vspace{0.5cm}

For this we will use the functional RG (FRG), which is defined with the Euclidean metric, and we focus 
more specifically on the Wetterich equation \cite{RiWe1990_1,RiWe1990_2,We1993_1,We1993_2,Mo1994_1,
Mo1994_2,internal2001_1,internal2001_2,internal2001_3}. For a single scalar field, this equation reads 
\begin{equation}\label{Wetterich}
k \partial_k \Gamma_k[\phi]
= \hf \mbox{Tr} \left[ \frac{k\partial_k R_k}{\Gamma^{(2)}_k[\phi]+R_k} \right]~,
\end{equation}
where $\Gamma_k[\phi]$ is the average effective action at the momentum scale $k$, and $R_k$ 
is an infrared regulator which freezes the infrared (IR) modes with momentum smaller than the scale $k$. 
A thorough account of the traditional and current applications of functional RG in various fields can be found 
in Ref.\,\cite{dupuis2021}.

In order to solve the previous equation we use the local potential approximation (LPA),
where the ansatz for the average effective action takes the form
\begin{equation}
\label{lopo_eff}
\Gamma_k[\phi]= \int {\mathrm d}^{d} x \left[\hf \partial_{\mu}\phi \partial^{\mu}\phi +  V_k(\phi)\right].
\end{equation}
and corresponds to the leading order expression for the action in the gradient expansion.
The Wetterich equation \eq{Wetterich} reduces then to a differential equation for the scale dependent 
potential $V_k(\phi)$  
\bea
\label{lpa_erg_pre}
k \partial_k {V_k(\phi)} 
=  \hf \int_{0}^{\infty}  \frac{d^d p}{(2\pi)^2} ~ \frac{k\partial_k R_k}{R_k + p^2 + V''_k}~,
\eea 
where $V''_k = \partial^2_\phi V_k$. An important comment is the following: the latter equation 
depends only on the second field derivative of the running potential, which allows to consider the 
singularity $|\phi|$. Indeed, 
\bea
\frac{\partial}{\partial\phi}\Big(V_k(|\phi|)\Big)&=&\mbox{sign}(\phi)V_k'(|\phi|)\\
\frac{\partial^2}{\partial\phi^2}\Big(V_k(|\phi|)\Big)&=&V_k''(|\phi|)~,\nonumber
\eea
and the equation \eq{lpa_erg_pre} is not sensitive to the absolute value of the field. 
An alternative argument is based on the regularisation introduced earlier
\be
V_{\mr{reg},k}(\phi)\equiv V_k\big(\sqrt{r^2+\phi^2}~\big)~,
\ee
from which one can note that
\be
\lim_{r\to0}V_{\mr{reg},k}''(\phi)=V_k''(|\phi|)~.
\ee

Let us demonstrate this general feature on a specific example. We consider the second 
derivative of the regularised and original dimensionful exponential potentials
\bea
\label{def_exp_dimful}
V_{\mr{EXP}} (\phi) &=& - u_k \exp(- a \vert \phi \vert) \nn
V_{\mr{EXP,reg}} (\phi) &=&  - u_k \exp(- a \sqrt{r^2 + \phi^2})
\eea
where the scale dependence is encoded in the amplitude $u_k$. Important to note that 
the dimensionful parameter $a$ is scale-independent in LPA since in this case the 
wavefunction renormalization $z$ is kept constant and by an appropriate rescaling of 
the field $\phi' \to a \phi$ these two couplings can be related to each other $z = 1/a^2$.
The second derivative of the regularised potential can be taken in the limit $r\to 0$ 
\begin{align}
&V''_{\mr{EXP,reg}} (\phi) =
-  u_k  \frac{a^2 \phi^2 \sqrt{r^2 + \phi^2} - r^2}{(r^2 + \phi^2)^{3/2}} \exp(- a \sqrt{r^2 + \phi^2}) \nn
&\lim_{r\to 0} V''_{\mr{EXP,reg}} (\phi) = -u_k a^2  \exp(- a \vert \phi \vert),
\end{align}
which is equivalent to the second derivative of the original non-analytic potential.
This indicates that the functional RG equation \eq{lpa_erg} is not sensitive to
the non-analytic nature of the potential.

Before studying more general non-differentiable potentials and the exact RG flow let us 
first discuss the linearised RG equations and the existence of topological phases.

\subsection{Linearised RG and topological phases}

One can write eq.(\ref{lpa_erg_pre}) as
\bea
\label{lpa_erg}
k \partial_k {V_k(\phi)} 
=  - \alpha_d k^d \int_{0}^{\infty}  dy  \, \frac{r' \, \, y^{\frac{d}{2}+1}}{[1+r] \, y \, + \frac{V''_k}{k^2}}\,,
\eea 
with $\alpha_d = \Omega_d/(2(2\pi)^d)$ where $\Omega_d = 2 \pi^{d/2}/\Gamma(d/2)$ 
and $r(y) = R/p^2$ is the dimensionless regulator with $y=p^2/k^2$ while $r' = dr/dy$. 

One can also introduce the dimensionless quantities, ${\tilde\phi} = k^{-{\frac{d-2}{2}}}\phi$, 
$V_k(\phi)=k^d {\tilde V_k({\tilde\phi})}$ where the corresponding dimensionless FRG
equation reads as
\bea
\label{lpa_erg_dimless}
\left(d-\frac{d-2}{2} \tilde \phi \partial_{\tilde\phi} + k \partial_k \right) \tilde V_k(\tilde\phi) =  \nn
- \alpha_d   \int_{0}^{\infty}  dy  \, \frac{r' \, \, y^{\frac{d}{2}+1}}{[1+r] \, y \, + \tilde V''_k}
\eea 
which is valid for the scale-dependent dimensionless potential with arbitrary regulator functions. 
One can apply further approximations, e.g, the linearisation of the FRG equation around the 
Gaussian fixed point where one finds 
\bea
\label{lpa_lin_d}
\left(d- \frac{d-2}{2} \tilde\phi \partial_{\tilde\phi} +k\partial_k \right) {\tilde V}_k(\tilde\phi) 
\approx - \alpha_d \,\, C \,\, {\tilde V''}_k(\tilde\phi)  
\eea
where the constant $C$ is usually regulator-dependent (except for $d=2$ where $C=1$ 
for any choice of the regulator function). One can use for example the Litim regulator  
\cite{Li2001_1,Li2001_2,Li2001_3} given by 
\be
\label{litim}
R_k(p^2)=(k^2-p^2)\Theta(k^2-p^2)~,
\ee
where one finds $C=2/d$ or the sharp-cutoff which gives $C=1$ in arbitrary 
dimension. In this section we use the sharp-cutoff regulator function. 

Let us first apply the linearised FRG equation \eq{lpa_lin_d} for the sine-Gordon
model \eq{sg} in $d=2$ dimensions which has the following potential, 
\beq
\label{def_sg}
\tilde V_{\mr{SG}} (\phi) = \tilde u_k \cos(\beta \phi)
\eeq
where the dimensionless Fourier amplitude carries the scale-dependence since in 
LPA the frequency $\beta$ does not depend on the running momentum cutoff $k$. 
The reason is that in LPA the wavefunction renormalization ($z$) is kept constant 
(it is not a running coupling) and this results in a constant frequency since 
$1/\beta^2 \equiv z$ which can be shown by rescaling the field variable. 
Let us note, the field carries no dimensions in $d=2$, thus, $\tilde \phi = \phi$ and
consequently $\tilde \beta = \beta$.
It is clear that the linearised FRG equation \eq{lpa_lin_d} preserves the functional 
form of the bare potential (no higher harmonics are generated) and one finds
\bea
\label{lin_sg}
(2+k\partial_k)\tilde u_k\cos(\beta\phi) =  \frac{1}{4\pi} \beta^2\tilde u_k\cos(\beta\phi)
\eea
and the RG flow equation for the Fourier amplitude reads
\bea
\label{lin_sg2}
k\partial_k\tilde u_k = \tilde u_k \left(-2+\frac{1}{4\pi}\beta^2 \right)
\eea
exhibiting the solution
\begin{equation}
\label{sg_linsol} 
{\tilde u}_k = {\tilde u}_\Lambda  \left(\frac{k}{\Lambda}\right)^{-2 + \frac{\beta^2}{4\pi}} 
\hskip 0.5cm \to \hskip 0.5 cm \beta_c^2 = 8\pi
\end{equation}
where ${\tilde u}_\Lambda$ is the initial (bare) value of the Fourier amplitude at the high 
energy ultra-violet (UV) cutoff $\Lambda$. Eq.~\eq{sg_linsol} determines the critical 
frequency $\beta_c^2 = 8\pi$ where the model undergoes a topological phase transition.
The coupling $\tilde u_k$ is irrelevant (decreasing) for $\beta^2>\beta^2_{c}$ and relevant 
(increasing) for $\beta^2<\beta^2_{c}$.

Now we turn to the non-differential potential \eq{exp_abs}, more precisely the one
defined in \eq{Sbranon}
\beq
\label{def_exp}
\tilde V_{\mr{EXP}} (\tilde \phi) = -\tilde u_k \exp(- \tilde a_k \vert \tilde \phi \vert).
\eeq
If one first considers \eq{def_exp} in $d=2$ dimensions, similarly to the SG model, 
the dimensionless amplitude $\tilde u_k$ carries the scale-dependence and the 
parameter $a_k = a$ does not depend on the running momentum cutoff $k$ in LPA.
In this case the linearised FRG equation gives,
\bea
\label{lin_exp}
(2+k\partial_k)\tilde u_k \exp(- a\vert \phi \vert) 
=  -\frac{1}{4\pi} a^2 \tilde u_k \exp(-a \vert \phi \vert)
\eea
and the RG flow equations for $\phi>0$ and for $\phi<0$ are identical and reads
\bea
\label{lin_exp2}
k\partial_k\tilde u_k = \tilde u_k \left(-2-\frac{1}{4\pi}a^2 \right)
\eea
which clearly shows the absence of any topological phase transitions (note
the sign change compared to the SG model). It is useful to compare our results with
those obtained for the Liouville model which depends on the field and not its absolute
value which are discussed in \cite{liouville}. One expects the same findings. Indeed,
the linearised RG flow equation \eq{lin_exp} is identical to Eq.~(28) of \cite{liouville}
(for $Q^2 = 1$) and the solution \eq{lin_exp2} is the same as Eq.~(29) of \cite{liouville}
(for $Q^2 = 1$). Note that the RG flow of the Liouville model (with or without 
the absolute value of the field) is identical to that of the so called sinh-Gordon model,
see \cite{shg}.

Let us now consider the $d\neq 2$ case where the linearised RG equation results in
\bea
\label{lin_exp_d}
\left(d - \frac{d-2}{2} \tilde\phi \partial_{\tilde\phi}  + k\partial_k \right)
 \tilde u_k   \exp(- \tilde a_k \vert \tilde \phi \vert) \nn
= - \alpha_d \tilde a_k^2 \tilde u_k   \exp(- \tilde a_k \vert \tilde \phi \vert)
\eea
where a scale-dependent (dimensionless) frequency $\tilde a_k$ should be 
introduced in order to keep the argument of the exponential term dimensionless. 
Thus, one expects a trivial scaling for the frequency, i.e., $\tilde a_k \sim k^{(d-2)/2}$.
In order to show this, one has to rewrite Eq.~\eq{lin_exp_d} for $\phi>0$ and for
$\phi<0$ and simultaneously one should separate the FRG equations into RG flow 
equations for $\vert \tilde \phi \vert \exp(- \tilde a_k \vert \tilde \phi \vert)$ and for  
$\exp(- \tilde a_k \vert \tilde \phi \vert)$,
\bea
\left(d  + k\partial_k \right) \tilde u_k  = - \alpha_d \tilde a_k^2 \tilde u_k    \nn
\pm \left(-\frac{d-2}{2} + k\partial_k\right) \tilde a_k = 0
\eea
where the sign $\pm$ of the left hand side of the last equation depends on
whether we consider the case $\phi>0$ or $\phi<0$, nevertheless, this 
difference does not matter because the right hand side of this equation is zero. 
Thus, the RG flow equations reads as
\bea
\label{lin_exp_d_sol}
&k\partial_k\tilde u_k = \tilde u_k \left(-d-\alpha_d \tilde a_k^2 \right), \nn
&k\partial_k \tilde a_k =  \frac{d-2}{2}  \tilde a_k
\hskip 0.5cm \to \hskip 0.5 cm  
 \tilde a^2_k =  \tilde a^2_\Lambda \left(\frac{k}{\Lambda}\right)^{d-2}
\eea
where the second RG flow equation gives back exactly the trivial scaling for the 
dimensionless frequency. We conclude that the RG study of the non-differentiable 
Liouville potential is identical to the that of the differentiable one, at least in the 
linearised regime. It is useful to rewrite \eq{lin_exp_d_sol} for dimensionfull
couplings, $u_k = k^d \tilde u_k$, $a = k^{2-d} \tilde a_k$,
\beq
\label{dimful_lin_exp_d_sol}
k\partial_k u_k = - k^{d-2} u_k \alpha_d a^2.
\eeq

Next, we focus on the solution of the linearised RG equation for a simple example 
of $V\propto|\phi|$ in order to confirm, as could be expected, that quantum fluctuations 
smoothen the bare potential and leads to a differentiable Wilsonian effective potential.

\subsection{Smoothening through quantum fluctuations}
We give here a simple illustration of how quantum fluctuations smoothen an non-differentiable 
microscopic potential. We start with the microscopic piecewise linear potential
\be\label{nondiffpot}
V_\infty(\phi)=\mu^{d/2+1}|\phi|~,
\ee
where $\mu>0$ is the only bare parameter of the model.
A linear potential does not get quantum corrections, so that we expect the IR effective potential $V_0(\phi)$ 
to be close to the bare potential (\ref{nondiffpot}) for $|\phi|>>\mu$, where the singularity is not felt. 
Hence we consider the natural ansatz, 
\be\label{ansatz}
V_k(\phi)=c_k+\mu^{d/2+1}|\phi|+u_k\exp(-|\phi|/\mu^{d/2-1})~,
\ee
where $u_k$ is to be determined and $c_k$ corresponds to a redefinition of the origin of energies, for each value of $k$. 

For the sake of simplicity we consider the cut-off function \eq{litim}, in which case eq.(\ref{lpa_erg_pre}) leads to
\be
k\partial_k V_k=\frac{2\alpha_d}{d}\frac{k^{d+2}}{k^2+V_k''}~.
\ee
Plugging the ansatz \eq{ansatz} in the latter evolution equation and ignoring field-independent terms gives
\be
ku_k'e^{-|\phi|/\mu^{d/2-1}}=\frac{2\alpha_d}{d}k^d\left(-\frac{u_k}{\mu^{d-2}k^2}e^{-|\phi|/\mu^{d/2-1}}+\cdots\right)~,\nonumber
\ee
where dots denote higher orders in the exponential. Dropping the higher order terms one finds,
\beq\label{ukprime}
ku_k' = -\frac{2\alpha_d}{d}\left(\frac{k}{\mu}\right)^{d-2} u_k~.
\eeq

\vspace{0.5cm}

\begin{center}
\begin{figure}[t]
	\includegraphics[scale=0.3]{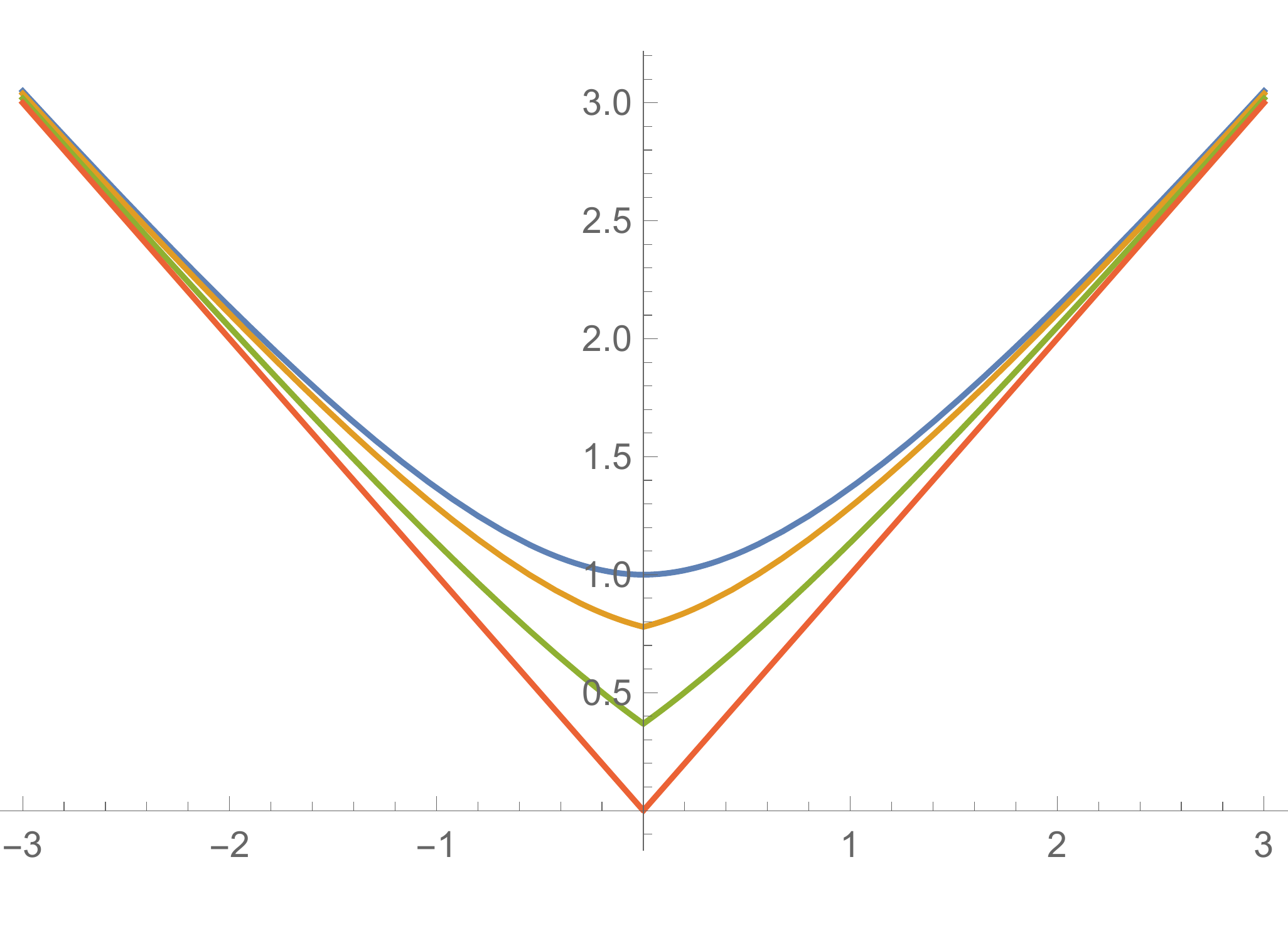}
	\caption{Quantum corrections turn the singularity into a smooth IR potential ($d=4$): the V-shaped 
	bare potential corresponds to $k=\infty$ and the smooth Wilsonian effective potential to $k=0$. The 
	intermediate running potentials correspond to finite values of $k$. Note that the latter potentials are 
	not differentiable, and only the deep IR effective potential is differentiable.}
	\label{Fig:nondiff}
\end{figure}
\end{center}

\noindent{\bf Case $d>2$}\\
In this case the solution of the latter equation reads
\be
u_k=A\exp\left(-\frac{2\alpha_d(k/\mu)^{d-2}}{d(d-2)}\right)~,
\ee
and the constant $A$ can be determined by imposing the IR potential $V_0(\phi)$ to be differentiable 
and minimum at $\phi=0$:
\be
V_0'(0)=\pm\left(\mu^{d/2+1}-\frac{A}{\mu^{d/2-1}}\right)=0~,
\ee
where the sign $\pm$ depends on which side of 0 the derivative is taken from. 
Hence the running average effective potential is
\bea\label{runningpot}
V_k(\phi)&=&c_k+\mu^{d/2+1}|\phi|\\
&&+\mu^d\exp\left(-\frac{2\alpha_d(k/\mu)^{d-2}}{d(d-2)}-\frac{|\phi|}{\mu^{d/2-1}}\right)~,\nonumber
\eea
and indeed corresponds to what is expected - see Fig.(\ref{Fig:nondiff}) for the case $d=4$:
\begin{itemize}
\item the ultraviolet limit $k\to\infty$ reproduces the bare potential;
\item the IR limit $k\to0$ leads a differentiable effective potential.
\end{itemize}

We note that Wilsonian renormalisation flows depend on the blocking procedure, but the IR effective 
potential obtained in the limit $k\to0$ does not. As an illustration we consider here the alternative 
Wegner-Houghton exact Wilsonian renormalisation equation \cite{WH}, which corresponds to the use 
of a sharp cut-off function, and which can be used in the local potential approximation. 
This equation is, up to field-independent terms,
\be
k\partial_k  U_k(\phi)=-\alpha_d k^d\ln\left(\frac{k^2+ U_k''(\phi)}{\mu^2}\right)~,
\ee
and the ansatz (\ref{ansatz}), followed by the steps described above, leads to
\bea
U_k(\phi)&=&\tilde c_k+\mu^{d/2+1}|\phi|\\
&&+\mu^d\exp\left(-\frac{\alpha_d(k/\mu)^{d-2}}{d-2}-\frac{|\phi|}{\mu^{d/2-1}}\right)~,\nonumber
\eea
where $\tilde c_k$ corresponds to a redefinition of the origin of the energies.
One notes the factor $\alpha_d$ instead of $2\alpha_d/d$ in the running potential (\ref{runningpot}), 
which is due to a different blocking procedure, but the IR effective potential is identical to the one 
obtained with the Wetterich equation and reads (for $c_0=\tilde c_0=0$)
\be
U_0(\phi)=V_0(\phi)=\mu^{d/2+1}|\phi|+\mu^d\exp\left(-\frac{|\phi|}{\mu^{d/2-1}}\right)~.
\ee

\vspace{0.5cm}

\noindent{\bf Case $d=2$}\\
In this case the differential equation (\ref{ukprime}) leads to
\be
u_k=u_\Lambda\left(\frac{\Lambda}{k}\right)^{1/4\pi}~,
\ee
where one recovers the anomalous dimension $1/4\pi$ appearing in Eq.~\eq{lin_exp2} for $a=1$.

\section{Functional RG equation for $O(N)$ symmetric scalar models}

In general the action of a scalar field theory which is symmetric under the $O(N)$ 
transformations where the wave-function renormalization is set to one, takes the form
\beq
\label{on_action}
S_{\rm ON}[\boldsymbol{\phi}]=\int \left[ \hf (\partial_\mu \boldsymbol{\phi})^2 
+ V_{\rm ON}(\boldsymbol{\phi})
\right]
\eeq
where $\boldsymbol{\phi}$ is a vector with $N$ components, and the potential 
depends on the magnitude of $\boldsymbol{\phi}$ due to the $O(N)$ symmetry. 
A new field variable $\rho=(1/2) \boldsymbol{\phi}^2$ can be introduced, and the 
FRG equation in LPA reads as
\bea
\label{on_frg_flow}
&&k \partial_k V_k = \\
&&\int_p \frac{k \partial_k R_k}{2} \left( \frac{1}{p^2+R_k+V'_k+2\rho V''_k} 
+ \frac{N-1}{p^2+R_k+V'_k}\right) \nonumber
\eea
where the prime means taking the derivative with respect to $\rho$. Note that one could also 
rewrite the flow equation in Eq.\,\eqref{on_frg_flow} in terms of the variable $\sqrt{2\rho}$ as it is 
done in Ref.\,\cite{yabu2017}. Using the Litim regulator \cite{Li2001_1,Li2001_2,Li2001_3} the 
momentum integral can be evaluated analytically and the following FRG equation can be derived 
for the dimensionless potential $\tilde V_k\equiv k^{-d} V_k$ 
\bea
\label{onrg}
k \partial_k \tilde V_k &=&(d-2) \rho \tilde V_k'-d \tilde V_k +\frac{(N-1)A_d}{1+\tilde V_k'}\\
&&~~~~~+\frac{A_d}{1+\tilde V_k'+2\rho \tilde V_k''}, \nn
A_d&=&\frac{1}{2^{d+1}}\frac{1}{\pi^{d/2}}\frac{1}{\Gamma(d/2)}\frac{4}{d}~,\nonumber
\eea
where $\tilde V_k'=\partial_{\rho}\tilde V_k$  and $\Gamma(x)$ is the gamma function.

Let us now consider the $O(N)$ extension of the potential \eq{abs},
\beq
\label{sqrtrho}
\tilde V_k(\rho)=\tilde u_k \sqrt \rho
\eeq
where, in contrast to the $N=1$ case, $\rho=\phi \cdot \phi/2$ is now the modulus of $N$ fields. 
Let us see whether the FRG approach can be applied and whether the non-differentiability at 
$\rho = 0$ presents a problem. Substituting this potential into Eq.~\eq{onrg} and neglecting 
field-independent terms yields
\beq
k \partial_k \tilde u_k \sqrt{\rho} =-(d/2+1) \tilde u_k \sqrt{\rho}+\frac{(N-1)A_d}{1+\tilde u_k/(2\sqrt{\rho})}.
\eeq
Since almost every term depends on the field $\sqrt{\rho}$ it is natural to expand this expression in 
the Taylor series of $\sqrt{\rho}$. This however, in case of $N>1$, generates higher powers of 
$\sqrt{\rho}$ which should be included in our ansatz \eq{sqrtrho}. Considering only the first order 
term, the renormalization scaling of the only coupling $\tilde u_k$ writes as
\beq
k \partial_k \tilde u_k= -(d/2+1) \tilde u_k + 2(N-1)A_d /\tilde u_k.
\eeq
For the dimensionful coupling the expression is simpler,
\beq
k \partial_k u=  k^{d+2} 2(N-1)A_d / u_k,
\eeq
with the solution,
\beq
u_k=
\sqrt{u_\Lambda^2 + \frac{4 (N-1) A_d }{d+2} (k^{d + 2} - \Lambda^{d + 2})},
\eeq
where the initial condition $u_{k=\Lambda}=u_\Lambda$ is given. It is clear, that when $k$ decreases 
$u_k$ decreases as well, however the running reaches a point at $u_{k_c}=0$ when it becomes unstable.
This can be avoided using $u_{k=0}=u_\mr{IR}$ as a boundary condition.
In this case, the coupling has the following scaling
\beq
u_k=
\sqrt{u_\mr{IR}^2 + \frac{4 (N-1) A_d}{d+2} k^{d + 2}},
\eeq
that can result a vanishing potential at $k=0$ if $u_\mr{IR}=0$ is chosen.

Thus, in the $O(N)$ extension of the V shaped potential, in contrast to the $N=1$ case, the coupling 
$u_k$ is running, but in the lowest order approximation, the initial condition can be chosen such that 
the potential and its non-analycitiy vanish. As noted before, the full analysis of this model would require 
higher order terms of $\sqrt{\rho}$, which are automatically generated by the RG running. However, the 
conclusion can be drawn, namely that the renormalization of \eq{sqrtrho} can be performed using the 
FRG approach without the need to resort to the regularizations presented in the potential \eq{abs_reg}.

Let us now consider the $O(N)$ extension of the SG model in $d=2$ dimensions where topological 
phase transition occurs for $N=1$.

The question is, again, whether one can apply the above FRG equation for the potential
 \beq
 \tilde V_k(\rho) = \tilde u_k \cos(\beta \sqrt{2} \sqrt{\rho}) 
 \eeq
which is $O(N)$ symmetric 
but also non-differentiable at $\rho = 0$. 
In two dimensions the FRG equation for $O(N)$ models reduces to
\bea
k \partial_k   \tilde V_k = -2 \tilde V_k + \frac{1}{4\pi} \frac{(N-1)}{1+\tilde V_k'} 
+ \frac{1}{4\pi} \frac{1}{1+\tilde V_k'+2\rho \tilde V_k''}.
\eea
and can be further simplified by linearising it around the Gaussian fixed point,
\bea
k \partial_k \tilde V_k \approx -2 \tilde V_k - \frac{N-1}{4\pi} \tilde V_k' -\frac{1}{4\pi} (\tilde V_k'+2\rho \tilde V_k'').
\eea
According to our previous arguments namely,
if the non-differentiability appears at a single point (single value for the field) the 
application of the usual FRG method is straightforward. Thus one finds,
\bea
k \partial_k   \tilde u_k \cos(\beta \sqrt{2} \sqrt{\rho}) \approx -2 \tilde u_k \cos(\beta \sqrt{2} \sqrt{\rho}) \nn
+ \frac{N-1}{4\pi}  \tilde u_k \beta \frac{\sin(\beta \sqrt{2} \sqrt{\rho})}{\sqrt{2}\sqrt{\rho}} 
+ \frac{1}{4\pi} \tilde u_k \beta^2 \cos(\beta \sqrt{2} \sqrt{\rho}),
\eea
and substituting back $\rho = \boldsymbol{\phi}^2/2$ the FRG equation reads 
\bea
k \partial_k   \tilde u_k \cos(\beta \sqrt{\boldsymbol{\phi}^2}) \approx -2 \tilde u_k \cos(\beta \sqrt{\boldsymbol{\phi}^2}) \nn
+ \frac{N-1}{4\pi}  \tilde u_k \beta \frac{\sin(\beta \sqrt{\boldsymbol{\phi}^2})}{\sqrt{\boldsymbol{\phi}^2}} 
+ \frac{1}{4\pi} \tilde u_k \beta^2 \cos(\beta \sqrt{\boldsymbol{\phi}^2}),
\eea
which contains periodic and non-periodic terms. For $N=1$, the non-periodic 
term vanishes and the flow equation for the dimensionless Fourier amplitude is
\bea
k \partial_k   \tilde u_k  \approx -2 u_k + \frac{1}{4\pi} \tilde u_k \beta^2,
\eea
which gives the well-known critical frequency $\beta^2_c = 8\pi$ of the single component 
SG model. This signals the presence of a topological type (infinite order) phase transition. 
For $N >1$, let us do the Fourier series of the non-periodic term keeping only a single mode.
One obtains
\bea
k \partial_k   \tilde u_k \cos(\beta \sqrt{\boldsymbol{\phi}^2}) \approx -2 \tilde u_k \cos(\beta \sqrt{\boldsymbol{\phi}^2}) \nn
+ \frac{N-1}{4\pi}  \tilde u_k \beta^2 \frac{Si(2\pi)}{\pi} 
\cos(\beta \sqrt{\boldsymbol{\phi}^2}) \nn
+ \frac{1}{4\pi} \tilde u_k \beta^2 \cos(\beta \sqrt{\boldsymbol{\phi}^2}),
\eea
where $Si(x)$ is the sine integral.
This gives the critical frequency
\beq
\beta^2_c(N)=\frac{8\pi}{1+ (N-1) Si(2\pi)/\pi}
\eeq
which is a decreasing function of $N$ and runs to zero in the large $N$ limit.
However, using the notation of \cite{O(N)_SG_2003}, rescaling the field and the frequency as 
$\hat\rho =( \boldsymbol{\phi} \cdot \boldsymbol{\phi} )/N$ and $\hat \beta = \beta \sqrt{N}$, 
one obtains a finite value in the large $N$ limit,
\beq
\hat \beta^2_c(N)=N \beta^2_c(N) \xrightarrow[]{N \to \infty}
\frac{8\pi^2}{Si(2\pi)}\approx 17.72 \pi \,.
\eeq
The value determined in \cite{O(N)_SG_2003} is $24\pi$. 

To have a better accuracy, the treatment of the non-periodic term would require a more careful 
analysis. Non-periodicity appears due to the fact that the Hessian in the FRG equation contains 
derivatives with respect to the Cartesian components of the field where an expansion around the 
field vector $\boldsymbol{\phi}=(\sqrt{2\rho},0,\cdots,0)$ was made to derive Eq.~\eq{onrg}. This 
is a sensible assumption for critical points where one expects an $O(N)$ symmetry breaking, as 
the expectation value of $\phi$ is directed only along a single direction, i.e. the first component. 
This assumption leads to the anisotropy between massive and Goldstone modes, which, in turns, 
causes the periodicity breaking. A possible solution is to perform the FRG analysis of the model in 
a spherical coordinate system, which deserves further studies.

Finally, we comment about the relation between these results and the well known 
Mermin-Wagner-Hohenberg-Coleman (MWHC) theorem \cite{Mermin66,Hohenberg67,Coleman73}: 
the latter applies to the $O(1)$ sine-Gordon model, and therefore when one finds a phase transition 
in it, one has to conclude that the transition is displaying spontaneous symmetry breaking (SSB) of 
a discrete symmetry (periodicity) not a continuous one, and we know that it is in the BKT class. One 
could do a similar argument for the $O(N)$ sine-Gordon model for any $N$, so that if one finds a 
phase transition in such model for $N>1$, one could conclude that the transition is not a SSB of the 
continuous $O(N)$ symmetry, but an SSB of the discrete symmetry, i.e., the periodicity. Indeed, we 
notice that the MWHC theorem prevents the SSB of the $O(N)$ symmetry, but not that of periodicity. 
One needs then to clarify what universality class the phase transition belongs to. So, if one can confirm 
(possibly by other tools) that at finite $N>1$ there is such a phase transition, it would be extremely 
interesting to classify what universality it is, what could be a lattice model realizing it, and what is the 
fate of such a transition for large $N$.

\section{Summary}

The main point of this article is to confirm an intuitive guess: quantum fluctuations smoothen a bare 
singularity. In quantum mechanics we have seen that the quantisation of a "regularised singularity" 
continuously leads to the result obtained without regularisation, when the regulator vanishes. In 
quantum field theory (QFT) we have seen that the Wilsonian running potential dresses the ultraviolet 
singularity along the renormalisation flow, to lead to a differentiable infrared potential. A more detailed 
analysis of the $O(N)$ SG model is necessary, but our results potentially open a new area of studies, 
allowing the treatment of singularities in a QFT, which could effectively arise from some high-energy 
models. 

It is worth noting that the smoothing of non-analytic potentials evidenced in our work appears to be a 
general property of the RG flow. This phenomenon has been linked to the entropy production that 
occurs during the RG flow, which, in turns, can be derived by relating Eq.\,\eqref{lpa_erg} to a non-linear 
advection diffusion process\,\cite{koenigstein2021II}, see also Refs.\,\cite{koenigstein2021I,koenigstein2021III}.

We note that the present QFT studies are based on the Wilsonian quantisation, and that the 
One-Particle Irreducible (1PI) quantisation of a singular potential might not be straightforward. Indeed, 
the usual perturbative expansions featured in the 1PI approach is based on Gaussian integrals, whereas 
the $V$-shaped potential would introduce error-functions instead, with the external source as the argument. 
A comparison between the two quantisation approaches would shed new light on the equivalence between 
the two effective potentials though, and is planned for a future work.

Finally, we observe that, since here we considered a finite number of isolated non-analytic points in which 
the singularity takes place, it would interesting to consider other types of non-differentiable potentials.

\section*{Acknowledgments}

Useful discussions with G. Somogyi are gratefully acknowledged. 
Support for the CNR/MTA Italy-Hungary 2019-2021 Joint Project 
"Strongly interacting systems in confined geometries" and funding 
from the Visegrad grant are gratefully acknowledged. The work of 
Jean Alexandre is supported by the STFC, UK, grant ST/P000258/1.

\appendix

\section{Effective Branon Action}
\label{app1}

To deal with a specific example of non-differentiable bare potential, we show here the construction 
of an effective description of brane fluctuations, described in the low energy limit by ``branons'', 
which are particles living in 3+1 dimensions, and representing quanta of the brane fluctuations in 
the 5th dimension.

Let us start with a general setup where a single brane model in large extra dimensions is 
considered. The four-dimensional space-time is embedded in the $D=4+N$ dimensional
bulk space.  In what follows, the brane coordinates are denoted with the indices $\mu,\nu$ 
and the bulk coordinates with $M,N$. In this general framework the coordinates parametrizing 
the points of in a bulk are denoted by $X^M=(x^\mu,y^m)$ and the position of the brane in
the bulk is given by $X^M=(x^\mu,Y^m(x))$, so thus $y^m = Y^m(x)$. Let us now switch back
to the simplest case, i.e., for $N=1$ and consider a 5-dimensional Universe with generic 
coordinates $X^M=(x^\mu,y)$, where $x$ are the coordinates on the brane, which is defined 
by the equation $y=Y(x)$.

Motivated by scenarios involving confinement on the brane, we consider the following 
block-diagonal bulk metric 
\be
g_{MN}=\left(\begin{array}{cc}e^{2\sigma(y)}\eta_{\mu\nu}&0\\0&-1\end{array}\right),
\ee
with the Randall-Sundrum warp factor, defined by
\be
\label{sigma}
\sigma(y)=-a|y|,
\ee
where $a$ is a constant. The induced metric $h_{\mu\nu}$ on the brane is 
\be
\label{induced}
h_{\mu\nu}(x)=\partial_\mu X^M\partial_\nu X^N g_{MN}(X)=e^{2\sigma(Y)}\eta_{\mu\nu}-\partial_\mu Y\partial_\nu Y,
\nonumber
\ee
and, if $f^4$ is the brane tension, the brane action is then 
\bea
S_{brane} =-f^4\int d^4x\sqrt{-h} = -f^4\int d^4x~e^{4\sigma(Y)} \\
\left( 1-\frac{1}{2}e^{-2\sigma(Y)}\eta^{\mu\nu}\partial_\mu Y\partial_\nu Y
+\cdots\right), \nonumber
\eea
where dots represent higher orders in derivatives of $Y$, which will be disregarded in the framework 
of the gradient expansion.\\
The dynamical variable is the canonically normalized branon field $\phi=f^2Y$, with mass dimension 1, 
and the classical brane ground state is $Y=0$. The resulting effective action for branons is then
\be
S_{branon}=\int d^4x\left( \frac{e^{2\sigma(\phi)}}{2}\partial^\mu\phi\partial_\mu\phi-f^4e^{4\sigma(\phi)}\right),
\ee
and contains derivative and polynomial interactions. For the details of the derivation see \cite{branon_stabilization}.

\section{Periodic Higgs Potential}
\label{app2}

In the standard model (SM) of particle physics the underlying symmetry of the electroweak sector is 
$SU(2)_L \times U(1)_Y$ and the Higgs Lagrangian reads as
\bea
&{\cal L} = (D_\mu \phi)^\star (D^\mu \phi) - V(\phi) - \frac{1}{2} \Tr \, ({F}_{\mu\nu} {F}^{\mu\nu}),  \nn
&D_\mu = \partial_\mu + i g {\bf T} \cdot {\bf W}_\mu +i g' y_j B_\mu   \nn
&\hskip -0.35cm V = \mu^2 \phi^\star \phi + \lambda (\phi^\star \phi)^2, \nonumber
\eea
where the SM Higgs field is an $SU(2)$ complex scalar doublet with four real components
\bea
\phi = \frac{1}{\sqrt{2}} \left(
\begin{array}{c}
\phi_1 +i \phi_2\\
\phi_3 +i \phi_4\\
\end{array}
\right). \nonumber
\eea
It is interesting to investigate the possibility of a false vacuum at large Higgs field values which can be realised by 
adding new interaction terms to the Lagrangian, such as $\lambda_2 (\phi^\star \phi)^3$. In the renormalization group (RG)
point of view the phase structure has not been modified significantly if the Higgs potential remains polynomial. Instead, 
a periodic self-interaction, \cite{sg_msg_higgs_1,sg_msg_higgs_2},
\bea
V = u[\cos(\beta \sqrt{\phi^\star \phi})-1] = -\frac{u \beta^2}{2}  \vert \phi^\star \phi \vert + \frac{u \beta^4}{4}   (\phi^\star \phi)^2 + ...
\nonumber
\eea
can influence more drastically the phase structure and the RG running of the couplings due to the periodicity. 
The first two terms of the Taylor expansion reproduce the SM Higgs potential but periodicity should be protected 
by the RG approach, so, no Taylor expansion can be applied.

Thus, it is worthwhile to study the phase structure of a periodic potential defined by the following Euclidean action
\begin{equation}
S[\boldsymbol{\phi}]=\int d^d x \,\Big( \hf \partial_{\mu} \phi_i \partial_{\mu} \phi_i 
+ u  \cos(\beta \sqrt{\boldsymbol{\phi} \cdot \boldsymbol{\phi}}) \Big)
\end{equation}
where summations over repeated indexes is intended. The index $i$ runs over the $N$ component of the field vector 
$\boldsymbol{\phi}(x)$, while $\mu$ is the spatial coordinate index. It is worth noting that the whole action is $O(N)$ 
invariant since the potential part depend only on the $O(N)$ invariant composition of the fields 
$\boldsymbol{\phi} \cdot \boldsymbol{\phi} =\phi_i(x) \phi_i(x)$.

Its phase structure has already been investigated for $d=2$ dimensions. 
For $N=1$, the scalar model \eq{o(N)_sg} reduces to the 2-dimensional 
sine-Gordon (SG) theory 
\begin{equation}
\label{sg}
S[\phi]=\int d^2 x \,\Big( \hf \partial_{\mu} \phi \partial_{\mu} \phi + u  \cos(\beta \phi) \Big)
\end{equation}
where we used the fact that cosine is an even function, so, its argument 
$\sqrt{\phi^2} = \vert\phi\vert$ can be replaced by $\phi$ dropping its absolute value. 
In this case the non-analytic nature plays no role and the 2-dimensional SG model 
undergoes a topological phase transition and its two phases are separated by the 
critical frequency 
\bea
\beta_{N=1}^2 = 8\pi. 
\eea
The $O(N)$ extension of the 2-dimensional SG scalar theory has been investigated in 
\cite{O(N)_SG_2003} with the following explicit form,
\begin{equation}
\label{andrea_o(N)_sg}
S[\phi] = 
\int d^2 x \,\Big( \hf \partial_{\mu} \boldsymbol{\phi}  \partial_{\mu} \boldsymbol{\phi} 
+ N \frac{\alpha_0}{\hat\beta^2}   \cos(\hat\beta \sqrt{\hat\rho}) \Big)
\end{equation}
where $\hat\rho =( \boldsymbol{\phi} \cdot \boldsymbol{\phi} )/N$ and $\hat \beta = \beta \sqrt{N}$. 
It was argued that similarly to the $N=1$ case, the general N-vector model has two phases and 
the critical frequency in the large N limit ($N\to \infty$) reads as
\beq
\hat\beta_{N\to\infty}^2=24\pi.
\eeq
However, the complete study of the phase structure of the general model, which is the $O(N)$ 
extension of the usual SG scalar theory is still missing.

\end{document}